# Elastic Cheerios effect: self-assembly of cylinders on a soft solid

ADITI CHAKRABARTI[1], LOUIS RYAN[2], MANOJ K. CHAUDHURY[1(a)] and L. MAHADEVAN[2,3(b)]

[1] *Department of Chemical and Biomolecular Engineering, Lehigh University, Bethlehem, PA 18015, USA*
[2] *Paulson School of Engineering and Applied Sciences, Harvard University, Cambridge, MA 02138, USA*
[3] *Department of Physics, Harvard University, Cambridge, MA 02138, USA*

**Abstract** – A rigid cylinder placed on a soft gel deforms its surface. When multiple cylinders are placed on the surface, they interact with each other via the topography of the deformed gel which serves as an energy landscape; as they move, the landscape changes which in turn changes their interaction. We use a combination of experiments, simple scaling estimates and numerical simulations to study the self-assembly of cylinders in this elastic analog of the Cheerios effect for capillary interactions on a fluid interface. Our results show that the effective two body interaction can be well described by an exponential attraction potential as a result of which the dynamics also show an exponential behavior with respect to the separation distance. When many cylinders are placed on the gel, the cylinders cluster together if they are not too far apart; otherwise their motion gets elastically arrested.

**Introduction.** – It is well-known that a small particle can float at an air-liquid interface due to the capillary force acting along its contact line [1-3]. The combination of the gravitational and the surface energies can lead to an attractive or a repulsive interaction between particles depending upon their specific gravity relative to the liquid [4-6]. This observation, dubbed the Cheerios effect [7-8] is the basis for capillarity driven self-assembly [9-12]. Similar phenomena are also observed on microscales in such instances as proteins embedded in a lipid membrane [13], in which the interactions are mediated by elasticity and capillarity. These observations lead to a natural question- what if the fluid interface is replaced by its elastic analog, such as the surface of a soft solid, or a thin elastic membrane? Recent experiments [14-16] have shown that this elastic analog of the Cheerios effect, wherein heavy spheres settling on a soft solid deform the interface and create a topography that serves as an energy landscape on which they move. These observations are consistent with a scaling theory [14-16] that captures the essential features for the forces and dynamics between two spheres. Here, we complement these studies by studying the statics and dynamics of heavy parallel cylinders sitting atop a soft substrate using a combination of experiments, theory and numerical simulations.

**Experiment.** – Our experiments followed a protocol similar to that in earlier studies [14-16] and used a physically cross-linked gel as the soft substrate, starting with a solution of N-(hydroxymethyl)-acrylamide (48% solution in water, Sigma Aldrich) in water that was then polymerized by adding 0.25wt% of the catalyst potassium persulfate (99.99%, Sigma Aldrich) and initiating the reaction with 0.3wt% *N,N,N',N'*- tetramethylethylenediamine (TEMED, $\geq$ 99.5%, Sigma Aldrich). After crosslinking was complete, we measured the shear modulus of this gel using an oscillatory rheology test [12] and found that the modulus $\mu$ = 18 Pa. For the cylinders, we used highly polished ¾" long aluminum rods (2024 Aluminum, 3/16 "diameter, density 2.8 g/cc, McMaster Carr) which were cleaned and sonicated in acetone and dried with ultrapure nitrogen gas. They were then plasma-oxidized and soaked in trimethylsiloxy-terminated polydimethylsiloxane (DMS T-22, M.W. 9430; Gelest Inc.), and baked at 80°C for a day to allow the polydimethylsiloxane chains to graft with the surface, and then rinsed with chloroform (ACS grade, EMD) in order to remove the unreacted siloxanes and dried.

When a single cylinder is placed on the gel surface, it deforms the interface locally. When another cylinder is placed within 5-8mm from the first one, they attract towards each other until they coalesce (Appendix, Movie 1) as shown in Fig. 1A. The interface was filmed with a CCD (charge coupled device) camera (MTI-72) that was equipped with a variable focal length microscope (Infinity), and the images analyzed using ImageJ to yield the surface of the gel ($h$)

(a)E-mail: mkc4@lehigh.edu
(b)Email: lm@seas.harvard.edu

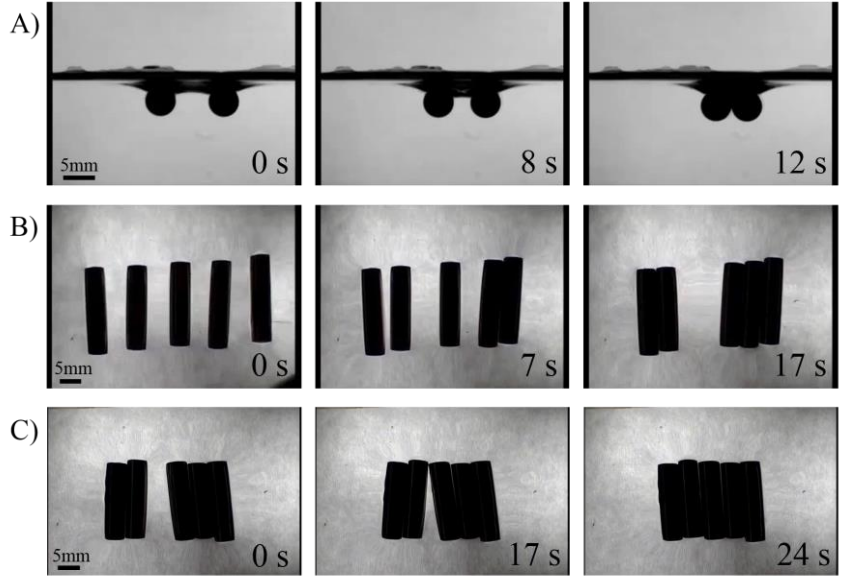

Fig. 1: (A) Two cylinders placed parallel to each other on a soft gel ($\mu$=18 Pa) move towards each other and eventually coalesce. (B) If the distance between the cylinders is large enough, we also see elastically arrested configurations as well. (C) A similar self-organized pattern to (A) arises for this initial configuration after sufficient time. The time stamps in the second and third images in the panel in C) are relative to the panel's first image.

as a function of the edge to edge separation distance ($\ell$) between them, as shown in Fig. 2C (inset). In Fig. 1B-C, we see that when there are more number of cylinders, they may undergo either complete or arrested coalescence (Appendix, Movies 2 and 3). In the experiments involving two or more cylinders, care was taken to keep them as parallel as possible.

To understand these results, we first consider the interaction of a slightly heavy cylinder of radius $R$ with a soft gel of shear modulus $\mu$. Assuming that the deformations of the relatively incompressible gel are small and of order $h$ in the vertical direction, the elastic energy of deformation of the medium per unit depth scales as $\mu \varepsilon^2 \ell^2$ where the strain $\varepsilon \sim h/R$, and $\ell$ is a characteristic horizontal scale over which the deformations decay. The gravitational energy of the deformed gel per unit depth scales as $\Delta \rho g h \ell^2$, where $\Delta \rho$ is the density difference between the cylinder and the gel. Balancing these energies yields the characteristic scale $h \sim \Delta \rho g R^2 / \mu$ over which deformations decay in the horizontal scale [17].

**Theory.** – A more formal analysis may be carried out by considering the total energy per unit width of the system, composed of the sum of the gravitational potential and elastic energy:

$$U = \left[\frac{\Delta \rho g}{2} \int_{-\infty}^{+\infty} \xi^2 dx + \frac{T}{2} \int_{-\infty}^{+\infty} \xi_x^2 dx\right] \\ + \left[\frac{\mu}{4} \int_{-\infty}^{+\infty}\int_{-\infty}^{0} (w_x + u_z)^2 dxdz + 2\mu \int_{-\infty}^{+\infty}\int_{-\infty}^{0} w_z^2 dxdz\right] \quad (1)$$

Where the free surface displacement is $\xi(x)$, the surface tension is $T$, and the vector of displacement fields in the incompressible solid of modulus $\mu$ is given by $(u(x,z), w(x,z))$, with $w(x,0) = \xi(x)$. Using a separable potential of the form $\chi(x,z) = \phi(x)\psi(z)$ to characterize the deformations with $u = \chi_z$ and $w = -\chi_x$, we see that the incompressibility condition $u_x + w_z = 0$ is automatically satisfied. Then, (1) may be rewritten as:

$$U = \left[\frac{\Delta \rho g}{2}\psi(0)^2\int_{-\infty}^{+\infty}\phi_x^2 dx + \frac{T}{2}\psi(0)^2\int_{-\infty}^{+\infty}\phi_{xx}^2 dx\right] \\ + \left[\frac{\mu}{4}\int_{-\infty}^{+\infty}\int_{-\infty}^{0}(\phi\psi_{zz} - \phi_{xx}\psi)^2 dxdz + 2\mu\int_{-\infty}^{+\infty}\int_{-\infty}^{0}\phi_x^2\psi_z^2 dxdz\right] \quad (2)$$

Functional minimization of $U$ with respect to $\phi$ along with the zero shear stress condition on the free surface ($\psi_{zz}(0)\phi(x) - \phi_{xx}(x)\psi(0) = 0$) leads to the following Euler-Lagrange equation:

$$\left(\frac{\mu}{2}\left(\int_{-\infty}^{0}\psi^2 dz\right) + T\psi(0)^2\right)\phi_{xxxx} \\ - \left[\Delta \rho \, g\psi(0)^2 + \mu\int_{-\infty}^{0}\psi_{zz}\psi dz + 4\mu\int_{-\infty}^{0}\psi_z^2 dz - \frac{\mu}{2}\int_{-\infty}^{0}\psi_{zz}^2 dz(\frac{\psi(0)}{\psi_{zz}(0)})\right]\phi_{xx} = 0 \quad (3)$$

For the case when the shear modulus vanishes, so that $\mu = 0$, the surface profile is controlled by the balance between tension and gravity, and we find that $\phi = \phi_0 e^{-\alpha x}$, and thus $\xi = \xi(0)e^{-\alpha x}$ with a decay length: $\alpha^{-1} = \sqrt{T/\Delta\rho g}$. In the case when the tension vanishes so that $T=0$, the surface profile is controlled simply by the balance between gravity and elasticity, and the surface profile is of the form $\xi = \xi(0)e^{-\beta x}$, where $\beta = \left[2\Delta\rho g\psi(0)^2 / \left(\mu\int_{-\infty}^{0}\psi^2 dz\right)\right]^{1/2}$ with $\beta^{-1} \sim \mu/\Delta\rho g$ being the decay length (we note that in this case the integrals of $\psi$ and its derivatives are all bounded). This implies that the descent of the cylinder can be expressed as $\xi_o \approx \Delta\rho g R^2/\mu$, which agrees with the scaling ansatz discussed previously. We emphasize that the functional minimization of $U$ has been carried out in the absence of the constraint: $\int_{-\infty}^{+\infty} \xi dx = 0$. Consideration of this condition in the constrained functional minimization only requires that all the measurements of $\xi_0$ need to be performed relative to the plateauing surface of the gel in the deformed state. The energetics remain unaltered; thus the final result of equation 3 remains the same.

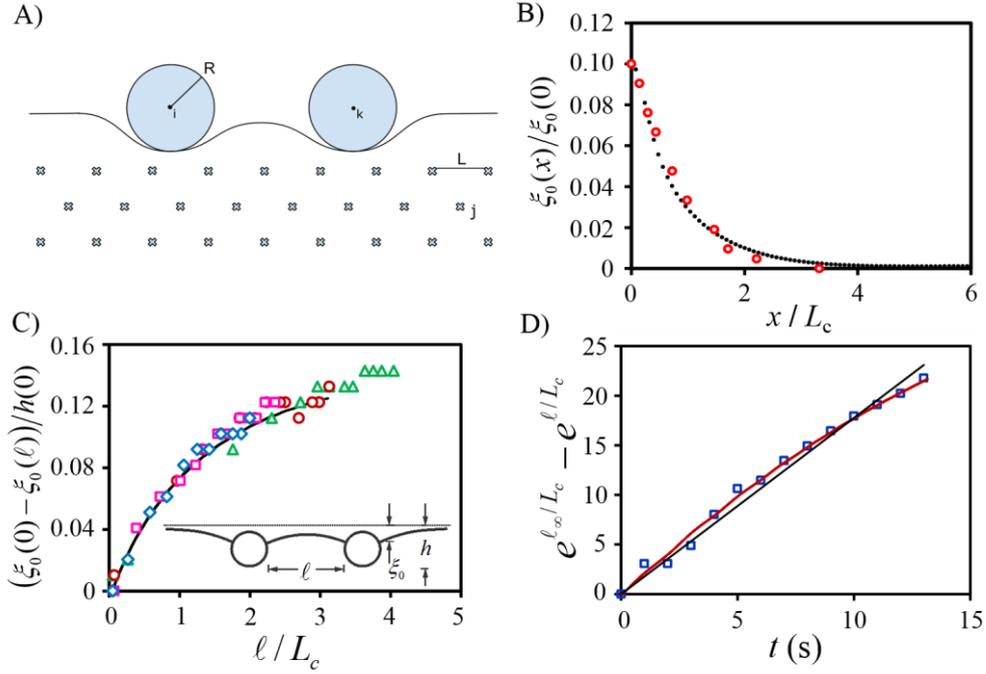

Fig. 2:: (A) A schematic illustrating the numerical model. (B) The deformed profile of the surface of a gel for a single cylinder. The red open circles show the experimental points and the black line are obtained from simulation. Both fit an exponential function $\xi_0(x) = \xi_o(0)\exp(-x/L_c)$, with $L_c$ = 7.23 (simulation) and 2.25 mm (experiment). (C) The settling depth of the cylinders $\xi_0(0) - \xi_o(\ell)$ scaled with their initial depth $h(0)$ is plotted as a function of the non-dimensional distance of separation $\ell/L_c$ where $L_c$ is the effective decay length of elastic deformations. The open symbols represent the data obtained from three different experiments. The black line shows the results of the numerical simulations of the equations of motion (4-6) with parameter values $R$ = 3, $K$ = 2000, $\rho$ = 0.35 with $\tau$ = 14.59. (Inset: Schematic of two cylinders approaching each other on the surface of a gel with appropriate notation used in the text) (D) The dynamics of attraction of two cylinders showing an exponential collapse (see text for details). The open symbols represent the experimental data. The red line is obtained from simulations. The black line corresponds to a linear fit of the experimental data.

**Simulation.** – To verify our scaling and analytic estimates, we now simulate numerically the interaction of the cylinders on a soft gel. Small particles are used to model the gel and large particles are used to model the cylinders that interact with the gel [18-19], as shown in the Fig. 2A. The equations of motion for the discretized gel are given by,

$$\ddot{x}_j + b\dot{x}_j = -\frac{\partial U}{\partial x_j}, \qquad (4)$$

$$U = \sum_{\bar{j}\in\text{Gel}} \frac{1}{2} K\left(\|x_j - x_{\bar{j}}\| - L\right)^2 \\ - \left(\frac{L}{\|x_j - x_{\bar{j}}\|}\right)^6 + \sum_{k\in\text{Cylinders}}\left(\frac{2R}{\|x_j - x_k\|}\right)^{12} \qquad (5)$$

Where we use a modified Lennard-Jones potential, that uses springs for short range repulsion, and the viscosity $b$ = 10 to eliminate any oscillations. The large particles representing the cylinders had the same equations of motion with the inclusion of a gravity term,

$$\ddot{x}_i + b\dot{x}_i = -\frac{\partial U}{\partial x_i} - \rho g R^2 \hat{y},$$

$$U = \sum_{j \in \text{Gels}} \left(\frac{2R}{\|x_i - x_j\|}\right)^{12} + \sum_{k \in \text{Cylinders}} \left(\frac{2R}{\|x_i - x_k\|}\right)^{12} \quad (6)$$

With $R = 3$, $g = 9.81$, $\rho = .35$. The equations (4-6) were numerically integrated using a leap-frog- integration scheme with a time step of $\Delta t = 0.01$, with the gel domain $[0, 100] \times [0, 40]$.

The results of our simulations confirmed the experimental observation that the profile of the deformed surface of the gel is indeed exponential with respect to $\ell$, i.e $\xi_0(\ell) = \xi_o(0)\exp(-\ell/L_c)$, as shown in Fig. 2B, upto a simple rescaling of $L_c$, and show that the omission of a logarithmic correction $\xi_0(\ell) \sim \ln(\ell/L_c)$ due to the classical Boussinesq stress field due to a line force exerted by the cylinder is justified. Simulations reproduced the general features of the attraction of two cylinders on a gel (Appendix, Movie 4) in that the experimental and theoretical results of the energy of two cylinders plotted as $[\xi_o(0) - \xi_o(\ell)]/h(0)$ versus $\ell/L_c$ exhibit excellent agreement with each other (Fig. 2C). Since the attractive energy of two parallel cylinders is exponential in $\ell$, the resulting force is also exponential. Assuming the friction between the cylinder and gel to be a linear function of their relative velocities, we can write: $-\partial U/\partial \ell \sim \zeta\, d\ell/dt$, $\zeta$ being the coefficient of kinematic friction. Integration of the preceding equation leads to $e^{\ell_\infty/L_c} - e^{\ell/L_c} \sim \left(\Delta\rho g R^2 L \xi_0(0)/\zeta L_c^2\right)t$, where $\ell_\infty$ is the initial distance of separation between the two cylinders. Plots of $e^{\ell_\infty/L_c} - e^{\ell/L_c}$ versus $t$ shown in Fig. 2D confirms the agreement between experiment and theory once the differences in kinematic friction are taken into account.

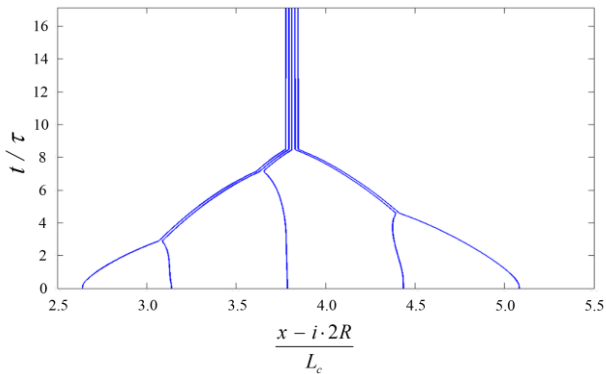

Fig. 3: The space time dynamics of coalescence of five cylinders. The variable *i* represents the index of the cylinder and is used to eliminate the space between cylinders in contact. These simulations solve the equations of motion given in (4-6), with the parameter values $R = 3$, $K = 500$, $\rho = 0.25$, yielding $L_c = 10.03$. The dynamics qualitatively capture the experimental scenario shown in Fig. 1, wherein the cylinders aggregate in pairs before slowing down and coalescing together.

Finally, in Fig. 3, we show the space-time plot of the coalescence of five cylinders as obtained from a numerical simulation, where we see that the rate of coalescence changes as the cylinders form pairs and then triplets, qualitatively consistent with our experimental observations (Fig.1B-C).

**Conclusion.** – Our study has uncovered the profile of the deformed gel around a single cylinder as a simple exponential that is consistent with a simple theory and corroborated by simulations, in which the gel was modelled as a network of beads connected and strings. When multiple cylinders are placed on the surface, they assemble to form smaller clusters that eventually aggregate to form one large cluster (Appendix, Movies 5 and 6), but also may be arrested elastically at times. Our elastic analog of the classical capillary attraction of particles at fluid interfaces opens a plethora of possibilities to be explored that are equally rich and perhaps more interesting, as elasticity provides an additional controllable degree of freedom.

∗∗∗

**Acknowledgment. -** We thank the NSF Graduate Fellowship Program (LR), the Harvard NSF-MRSEC DMR0820484 (LM) and the MacArthur Foundation (LM) for partial support.

**Appendix.** Six movies.

Movie 1: Coalescence of two rigid cylinders on gel ($\mu = 18$ Pa) (experiment).
Movie 2: Arrested coalescence of multiple cylinders (experiment).
Movie 3: Complete coalescence of multiple cylinders (experiment).
Movie 4: Coalescence of two cylinders on gel (simulation).
Movie 5: Arrested coalescence of multiple cylinders (simulation).
Movie 6: Complete coalescence of multiple cylinders (simulation).

REFERENCES

[1] ALLAIN C. and CLOITRE M., *J. Colloid Interface Sci.*, **157** (1993) 261.
[2] KRALCHEVSKY P. A. and NAGAYAMA K., *Adv. Colloid Interface Sci.*, **85** (2000) 145.
[3] CHAN D. Y. C., HENRY J. D. and WHITE L. R., *J. Colloid Interface Sci.*, **79** (1981) 410.
[4] CHAUDHURY M. K., WEAVER T., HUI C. Y. and KRAMER E., *J. Appl. Phys.*, **80** (1996) 30.
[5] NICOLSON M. M., *Proc. Cambridge Philos. Soc.*, **45** (1949) 288.

[6] BOTTO L., LEWANDOWSKI E. P., CAVALLARO M. and STEBE K. J., *Soft Matter,* **8** (2012) 9957.
[7] VELLA D. and MAHADEVAN L., *Am. J. Phys.*, **73** (2005) 817.
[8] LARMOUR I. A., SAUNDERS G. C. and BELL S. E., *Angew. Chem. Int. Ed*., **47** (2008) 5043.
[9] WHITESIDES G. M. and GRZYBOWSKI B., *Science*, **295** (2002) 2418.
[10] GRZYBOWSKI B. A., WILMER C. E., KIM J., BROWNE K. P. and BISHOP K. J., *Soft Matter,* **5** (2009) 1110.
[11] CAVALLARO M., BOTTO L., LEWANDOWSKI E. P., WANG M. and STEBE K. J., *Proc. Nat. Acad. Sci*., **108** (2011) 20923.
[12] SRINIVASAN U., LIEPMANN D. and HOWE R. T., *J. Microelectromech. Sys.*, **10** (2001) 17.
[13] KIM K., NEU J. and OSTER G., *Biophys. J.*, **75** (1998) 2274.
[14] CHAKRABARTI A. and CHAUDHURY M. K., *Langmuir*, **29** (2013) 15543.
[15] CHAKRABARTI A. and CHAUDHURY M. K., *Langmuir*, **30** (2014) 4684.
[16] CHAKRABARTI A. and CHAUDHURY M. K., *Langmuir*, **31** (2015) 1911.
[17] MORA S. and POMEAU Y., *arXiv preprint, arXiv:1405.0315*. (2014).
[18] HAILE J. M., *Molecular Dynamics Simulation: Elementary Methods* (John Wiley & Sons, Inc., New York) 1992.
[19] RAPAPORT D. C., *The Art of Molecular Dynamics Simulation* (Cambridge University Press) 2004.